\documentclass{article}
\usepackage{spconf,amsmath,graphicx,hyperref,cite}

\title{Detection and Statistical Modeling of Birth-Death Anomaly}
\name{Salman Mohamadi, Farhang Yeganegi, Nasser M Nasrabadi}
\address{Department of Electrical Engineering, West Virginia University, WV, USA\\ Department of Electrical Engineering, Amirkabir University of Technology, Tehran, Iran\\
	\tt sm0224@mix.wvu.edu, farhang.yeganegi@aut.ac.ir, nasser.nasrabadi@mail.wvu.edu}
\begin{document}
	\ninept
	\maketitle
	\begin{abstract}
		Generally, anomaly detection has a great importance particularly in applied statistical signal processing. Here we provide a general framework in order to detect anomaly through the statistical modeling. In this paper, it is assumed that a signal is corrupted by noise whose variance follows an ARMA model. The assumption on the signal is further compromised to encompass the inherent nonstationarity associated with natural phenomenon, hence, the signal of interest is assumed to follow an ARIMA model and the noise to denote an anomaly, however, unknown. Anomaly is assumed to possess heteroskedastic properties, therefore, ARCH/GARCH modeling could extract the anomaly pattern given an additive model for signal of interest and anomaly.
	\end{abstract}
	\section{Introduction}
	\label{sec:intro}
	
	\subsection{Anomaly}
	\label{ssec:subhead}
	
	Anomaly or outlier detection \cite{abraham1989outlier, chandola2009anomaly}, is a technique that can detect the presence of abnormal signals. Anomaly is a pattern in the data that does not conform to the normal state or behavior. Important applications include the detection of radar and sonar system intruder, covert telecommunication, cyber intrusion, biomedical anomalies. In signal processing applications, detection of unknown targets (signals) are achieved adaptively using a CFAR (constant false alarm rate)-based technique has been widely used to resolve detection for many problems for other applications. Anomaly detection can be achieved by identifying instances signals, objects, and situations that deviate from the expected, known or {\em normal} behavior and thus may be of interest for further investigation. Analytically, the methods that have been proposed and implemented for anomaly detection  are based on statistical modeling of the anomalous signals, objects, \ldots, etc. The existing algorithms for anomaly detection are data driven; that is, normalcy is determined by machine learning algorithms analyzing a relative large set of historical data assumed to reflect normalcy. Generally, the methods are based on either neural networks that learn what is normal by unsupervised/semi-supervised learning, or more refined and transparent statistical/probabilistic models, or a hybrid approach where neural networks are used for determining parameters of a statistical model. Considering the statistical methods, these can be categorized as parametric or non-parametric, where the parametric methods assume that the model for (normal) data has a particular structure or belongs to a family of parameterized models. Structure and parameter setting can be purely data driven, e.g. unsupervised learning of structure and parameter estimation based on available data using machine learning techniques, or it can be a hybrid approach supporting the incorporation of human expert knowledge together with unsupervised/supervised learning. For example, techniques for reliably detecting and precisely time localizing an intruder in a radar system, or the onset of a disease such as seizure attacks are important for biomedical practitioners to reduce possible heart failure, to detect a seizure, it is necessary to monitor many ECG/EEG signals from a patient at the same time which leads to a large-scale inference problem involving either producing massive amount of real-time estimates or testing hypotheses at each instance of time.
	We propose to model the measured signal comprised of two components, a normal state denoted by ($N$) with no anomaly present, and an anomalous state having described by the normal state; i.e. normal signal, in addition to anomalous signal; i.e. anomalous state denoted by (A). The task is to investigate whether in the data there exists an anomalous state (signal). This application gives rise to the anomaly or signal detection problem, which can be stated as a hypothesis testing problem. Generally speaking, normal state is expected to exist for all times, and for some time anomalous signal coexists with the normal state or to exist on its own. Hence, we could face the birth and death of normal and/or anomalous state as shown in Figure \ref{Fig1bin}. In the rest of the paper, problem formulation in terms of time series modeling is provided, next application of adaptive CFAR detector for anomaly detection, and some simulations to validate the proposed approach, and we end the paper, with some concluding remarks.
	\subsection{Anomaly vesus heteroskedasticity}
	\label{ssec:subhead}
	In a general point of view, mostly, anomalous patterns in a 1-D time series would fall into four categories: Additive Anomaly (AA), Innovation Anomaly (IA), Level Shift Anomaly (LSA) and Transitory Change Anomaly (TCA). In this sense, we are intended to address the question that whether all type of anomalies in a 1-D time series, show properties of heteroskedasticity or some of them may arise of only homoscedastic properties. First we think of nature of these four type of anomalous pattern. Briefly to investigate, we have\\
	A) Additive Anomaly (AA): This type of anomaly affcts a single observation. After this disturbance, the series returns to its normal pattern as if nothing has happened. The effect caused by AA at a time t = T, with the magnitude of the effect is denoted by $\omega $ is given by
	$Z_t = X_t + \omega I_t^T = \frac{{\theta (B)}}{{\phi (B)}}{a_t} + \omega I_t^T$
	where: $I_t^T = 1, t=T$ and $I_t^T = 0, \rm{t} \ne \rm{T}$\\
	B) Innovation Anomaly (IA): It is the type of anomaly that affects the subsequent observations starting from its position or an initial shock that propagates in the subsequent observations. The effect of an IA is given by:
	$Z_t = X_t + \omega I_t^T = \frac{\theta (B)}{\phi (B)}({a_t} + \omega I_t^T)$
	
	An AA affects only the T observation, whereas an IA affects all observations beyond time T through the memory of the system described by $\frac{\theta (B)}{\phi (B)}$.\\
	C) Level Shift Anomaly (LSA): It emerges like a step function. For a stationary process, a level shift implies a change in the mean of the process after a point and consequently the process is turned into a non-stationary process.\\
	D) Transitory Change Anomaly (TCA): Transitory change anomaly (TCA) is a spike that fades exponentially after a few periods. The impact of a TCA is not permanent however it disappears exponentially.\\
	All of these types of anomaly contain the changes in amplitude though variance in the advent of and during the anomalous pattern, so some heteroskedastic features could be detected during the presence of anomaly \cite{engel1996forward,gonzalez2002combining,garcia2009anomaly,steinwart2005classification} 
	\section{Problem formulation}
	Many stochastic processes, in particular natural processes can be modeled as a birth-death processes, such processes are considered as a continuous-time Markov chains \cite{crawford2012transition}. In processes that show anomaly pattern especially natural processes, for instance, in epilepsy, even patients using drug to control the seizure, it occurs time to time \cite{adeli2007wavelet}. In formulating the simultaneous existence of normal and anomalous signal in the data set $S_t$, we assume the normal state is modeled by an ARIMA process $X_t,$ and the anomalous state by a GARCH (Generalized Autoregressive Conditional Heteroskedasticity) model $Z_t$, furthermore, in order to contemplate a unique equation for the binary hypothesis testing, a binary random process $a_t$ is assumed to be multiplied by $Z_t$ as follows
	\begin{equation}
	S_t=X_t+a_tZ_t
	\label{eqst}
	\end{equation}
	here $a_t$ is a binary process, which refers to the occurrences of birth-death anomaly and $X_t$ presents the main pattern of signal modeled by ARIMA and $Z_t$ is an anomaly pattern modeled by GARCH process. Main task is oriented toward the best model assignment individually to each part of $S_t$ in (\ref{eqst}). We assume $X_t$ is an existing {\em almost} stationary process modeled by an ARIMA model, therefore ARIMA at first would be fitted to model, but as already discussed, anomalous $Z_t,$ a nonstationary process, which exhibits heteroskedasticity in a birth-death alternative. GARCH is selected to catch this feature of variation of $S_t$. Orders of hybrid model ARIMA-GARCH are estimated using  AIC principle.
	\begin{center}
		\begin{figure}[h]
			\vspace{-4.5cm}
			\hspace{-.85cm}
			\includegraphics[width=11cm,height=11cm]{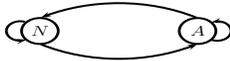}
			\vspace{-5.75cm}
			\caption{Birth-Death Process.}
			\label{Fig1bin}
			\vspace{-.5cm}
		\end{figure}
	\end{center}
	\begin{center}
		\begin{figure}[t]
			\includegraphics[scale=.5]{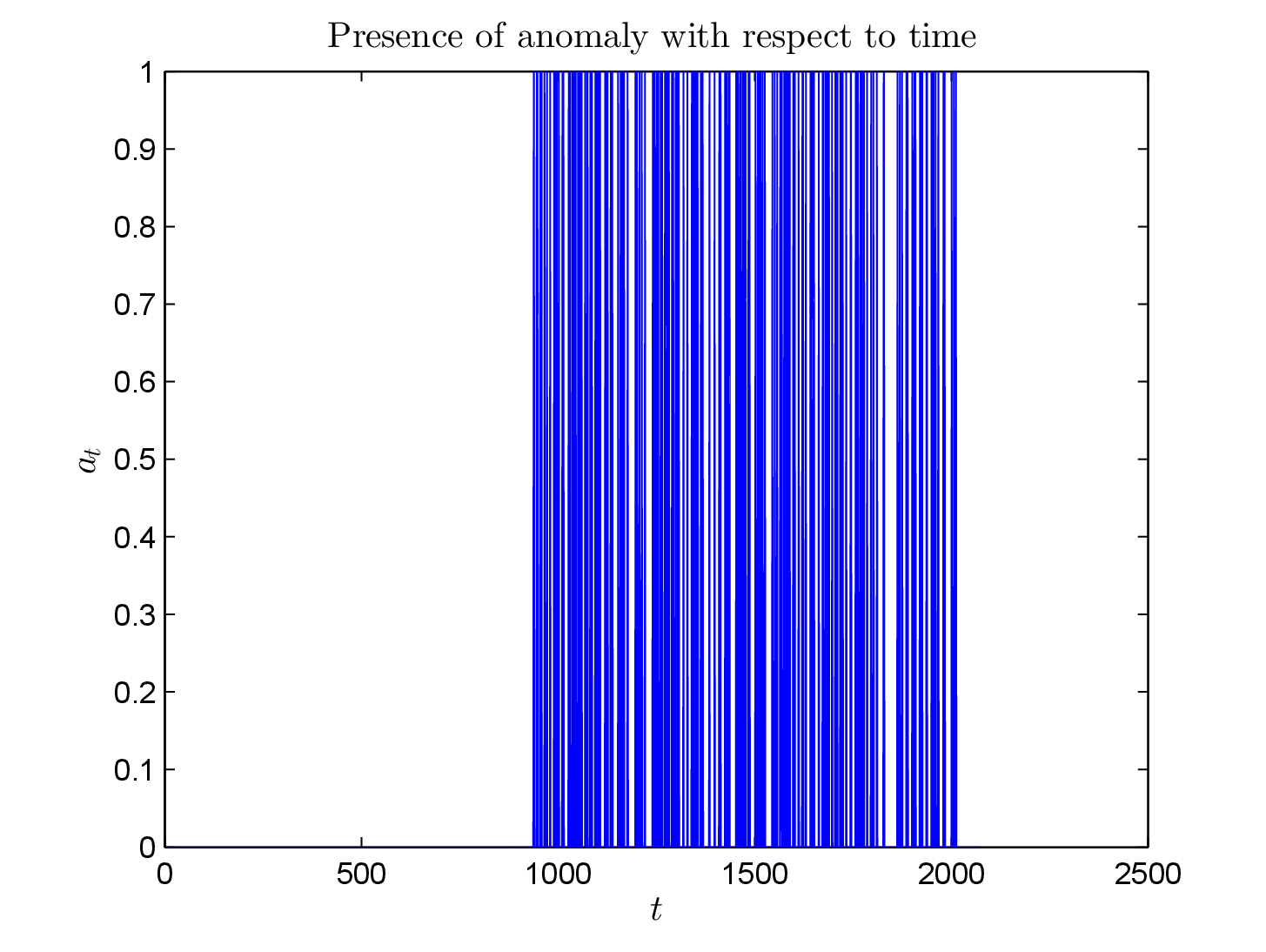}
			\caption{The presence of anomaly $Z_t$ with respect to time, the blue lines denote the weight of the times that anomaly is present.}
			\label{Fig2aoft}
		\end{figure}
	\end{center}
	As mentioned Figure~\ref{Fig1bin} depicts the birth-death scenario of normal and anomalous states in which the probability density function of the process $a_t$, $p(a;t)=(1-p_0)\delta(a_t=1)+p_0\delta(a_t)$ in (\ref{eqst}) that models the presence of an anomaly; i.e. $a_t=1$ or its absence $a_t=0,$ on the other hand, when $a_t=1$ with probability $1-p_0$ the anomaly denoted by $Z_t$ is invoked. A typical process $a_t$ is demonstrated in Figure~\ref{Fig2aoft}.
	Patterns denoted as anomaly pattern, may include a wide range of patterns violating the general trends of signal, statistical modeling here, plays an undeniable role, particularly in understanding the anomalies through modeling \cite{song2007conditional}. AR modeling and more effectively, ARMA and ARIMA, have proven to extract the existence of statistical pattern beneath the surface of visible anomalous fluctuations of time series \cite{pena2013anomaly,isbister2015anomaly}. On the other hand, while in many works, it can be assumed that the observed signal follows a Gaussian distribution, realistic natural signals well known to follow non-Gaussian distribution with time-varying variance. ARCH or GARCH, (Autoregressive conditional heteroskedasticity, Generalized autoregressive conditional heteroskedasticity) can capture the heteroskedasticity and non-Guassianity of the phenomenon of an anomaly. The time series exhibiting birth-death of an anomaly can be designed to possess a time varying conditional mean captured through the ARIMA modeling and a possible time varying volatility (standard deviation) expressed through the GARCH model, the combined hybrid model is termed as ARIMA-GARCH.
	\subsection{ARIMA--GARCH Modeling}
	ARIMA approach developed by Box and Jenkins \cite{box2015time} is a class of stochastic models used to analyze time series data. Consider the following autoregressive moving average model denoted as ARMA$(p, q)$	
	\begin{equation}
	\label{arma}
	S_t=\mu+\sum_{i=1}^p \phi_i S_{t-i}+\sum_{j=1}^q \theta_j Z_{t-j}+Z_t
	\end{equation}
	where $\mu$ is a constant term, $\phi_i$ the $i$th autoregressive coefficient, $\theta_j$ the $j$th
	moving average coefficient, and $Z_t$ the error term; usually zero mean Gaussian sequence, at time $t$, and $p$ and $q$ are called the orders of autoregressive and moving average terms respectively. When the backshift operator $B$ is applied, then, (\ref{arma}) is written as
	\begin{equation}
	\label{arima}
	(1-\sum_{i=1}^p\phi_i B^i)(S_t-\mu)=(1+\sum_{j=1}^q \theta_j B^j) Z_t
	\end{equation}
	where $B(S_t-\mu)=S_{t-1}-\mu$ and $BZ_t=Z_{t-1}$, the unknown parameters of this model by observing the time series $S_t$ can be obtained by a maximum likelihood estimation. In the traditional ARIMA models, error term $Z_t$ has zero mean and is homoscedastic plus the serial uncorrelated property. When the time series data exhibits the conditional heteroskedasticity, ARCH models proposed by Engle \cite{engle1982autoregressive} and later Bollerslev \cite{bollerslev1986generalized} are more appropriate and should be adopted. ARCH/GARCH models can accommodate the serial correlation in volatilities
	which changes over time. In an ARCH model, $Z_t$ is transformed as
	\begin{equation}\label{ARCH} Z_t=\sqrt{\nu_t} \varepsilon_t \end{equation}
	where $\varepsilon_t$ is an IID zero mean, unit variance Gaussian process, and $\mbox{Variance}(Z_t|\psi_{t-1})=\nu_t,$ where $\psi_{t-1}$ represents the information known before time $t$. Assume that $\nu_t$ is dependent on $\ell$ previous errors and can be estimated by the following equation,
	\begin{equation} \label{ARCH2}\nu_t=\varsigma_0+\sum_{i=1}^{\ell}\eta_i Z_{t-i}^2\end{equation}
	where $\varsigma_0$ and $\eta_i$ are constant coefficients, in this case $Z_t$ is said to follow an ARCH process of order $\ell,$ expressed as ARCH$(\ell)$. GARCH models are a generalized version of ARCH models and developed by Bollerslev \cite{bollerslev1986generalized}. In a GARCH model, the current conditional variance depends not only on $\ell$ previous errors but also on $k$ previous conditional variances. That is, equation (\ref{ARCH2}) becomes the following form, $Z_t$ is transformed as
	\begin{equation} \label{GARCH}\nu_t=\varsigma_0+\sum_{i=1}^k \varsigma_i \nu_{t-i}+\sum_{i=1}^{\ell}\eta_i Z_{t-i}^2\end{equation}
	in this case $Z_t$ is said to follow an GARCH process of order $(k,\ell)$.
	In the proposed method, in order to make the time series approximately uncorrelated, at first we use one order of differencing on $X_t$, i.e. $d_t = X_t - X_{t-1};$ then perform our modeling on $d_t$, ARIMA modeling, then, GARCH will be performed on ARIMA extracted residuals. AIC as one of best criterions, plays the important role for the order selection, in both ARIMA and ARCH/GARCH order selection. Referring back to equation~(\ref{eqst}), our aim is to detect the presence of $Z_t$, however, $X_t,$ the conditional mean of $S_t$ is ideally a stationary process. Signal detection in a nonstationary noise environment has been described extensively in the literature. In the derivation of a desired test procedure, one usually assumes either the model of a known deterministic signal in unknown correlated noise \cite{kelly1986adaptive}, or the model of an unknown random signal in white noise \cite{nitzberg1973constant}. The solution proposed for the former case is an adaptive implementation of
	the matched filter or a generalized likelihood ratio test, while that for the latter is the normalization technique. Furthermore, the detection of unknown random signals in unknown correlated noise has been adequately described in \cite{zhang1996cfar}. In this paper, $Z_t$ is a nonstationary signal that maybe present in the observation of $S_t$ in equation~(\ref{eqst}), on the recently, on the other hand, small target detection in the nonstationary environment has been investigated using stationary wavelet transform \cite{duk2017target}, this approach does not exactly correspond to equation (\ref{eqst}) because $a_t$ could cause the the anomaly to disappear for some time. We alleviate this phenomenon by adopting windows around each time slot of $S_t$ using adaptive threshold setting through CFAR scheme.
	\subsection{Anomaly  detection using CFAR detector}
	Presence or the absence of anomaly can be expounded through the application of detection theory which is a statistical tool used for decision making, this tool has been extensively studied and largely used
	in many fields, including: CDMA multiuser detection and pseudonoise code acquisition, OFDM signal detection, acquisition of weak GPS signals, spectrum sensing in cognitive radio, mobile localization, UWB localization for trough the wall imaging, distributed detection in sensor networks, adaptive subspace detection, failure detection in dynamic systems, target recognition for automotive applications. For a fixed and completely controlled false alarm rate, the CFAR-based detectors tries to do their best for maximizing the detection probability under a constant false alarm probability constraint. By doing that, the intention is focused at first on keeping the first type of detection errors under control; i.e. probability of false alarm, while the second type of detection errors; i.e. probability of a miss, is minimized at the best. In modeling the presence of anomaly through equation (\ref{eqst}) and ARIMA-GARCH modeling of $S_t$ to identify $X_t$ and $Z_t,$ the ARIMA, and GARCH components, respectively. Therefore, the detection problem for presence of anomaly is as follows
	\begin{equation} 
	\label{det}
	S_t=
	\left \{ 
	\begin{array}{l} X_t, \quad \mbox{no anomaly}\\ 
	X_t+Z_t \quad \mbox{anomaly is present.} 
	\end{array} 
	\right.
	\end{equation}
	In (\ref{det}), $X_t$ is an ARIMA$(p,q)$, and $Z_t$ is GARCH$(k,\ell)$, identified by the maximum likelihood algorithm \cite{box2015time}. We are encountered a nonstationary signal $Z_t$ whose standard deviation is time varying as described by equation (\ref{GARCH}), the probability density functions of $X_t$ is zero mean Gaussian, with constant variance, denoted by $p(X_t)$ and that of $Z_t$ is still zero mean, however, with time varying variance (determined by (\ref{GARCH})), denoted by $p(Z_t)$ as follows
	\begin{eqnarray} 
	\label{pdfs}
	p(X_t)&=&\frac{1}{\sqrt{2\pi \sigma^2_X}}\exp\left(-\frac{Z_t^2}{2\sigma^2_X}\right),
	\\
	p(Z_t)&=&\frac{1}{\sqrt{2\pi \nu_t}}\exp\left(-\frac{Z_t^2}{2\nu_t}\right)
	\end{eqnarray}
	In an EEG/ECG recording, $S_t$ is stochastic in nature and is attributed to its system dependencies; i.e. patient dependent. The general idea is to compare the squared amplitudes of the received volatility; i.e. the estimated time varying standard deviation of $Z_t$ to determine the presence of the anomaly. By windowing the time varying standard deviation of $Z_t$, $\nu_t$ and estimating locally a threshold using adaptive CFAR, hence, we monitor a rapid rise in the threshold locally, therefore, the presence of an anomaly is detected. For binary hypothesis testing in (\ref{det}), the probability density functions in (\ref{pdfs}) should be used and as it can be seen due to time varying nature of standard deviation (volatility) the threshold cannot be rendered as a constant, it time varying, this requires adaptive threshold setting through CFAR. In a typical adaptive detection scheme, the power of estimated volatility from (\ref{GARCH}) is used within the reference window composed of $2T$ reference cells. The guard cells, immediate neighbors of the cell under investigation (CUT), are excluded from the estimation process to avoid an eventual spillover from the CUI. An scaling factor is obtained via an optimization process where the criterion for it is a function of the probabilities of the two erroneous decisions, i.e. probability of miss and false alarm probability, for a window size comeon of ${t-T < t < t+T}$, the adaptive threshold setting uses
	\begin{equation} 
	\mbox{Variance}\left\{S_t\right\}=\left\{ 
	\begin{array}{l} \sigma^2_X+\xi_t, \quad \mbox{no anomaly}\\ 
	\nu_t+\xi_t, \quad \mbox{anomaly is present.} 
	\end{array} 
	\right.
	\end{equation}
	In this equation, $\mbox{Variance}\left\{S_t\right\}$ is the estimated vector of volatility (standard deviation) of observation signal, $S_t$ at time $t$ with a window size of length $2T$, the elements of this vector which can assume only noise values $\xi_t$, hence, no anomaly, or are greater than a time varying threshold $\nu_t$, therefore the presence of is announced. Due to impulsive nature of anomaly, there are rapid changes in the estimated $\nu_t,$ therefore, for nonhomogeneous $2T$ reference cells, order statistics CFAR yields better results.
	\begin{figure}[t]
		\centering{
			\includegraphics[scale=.4]{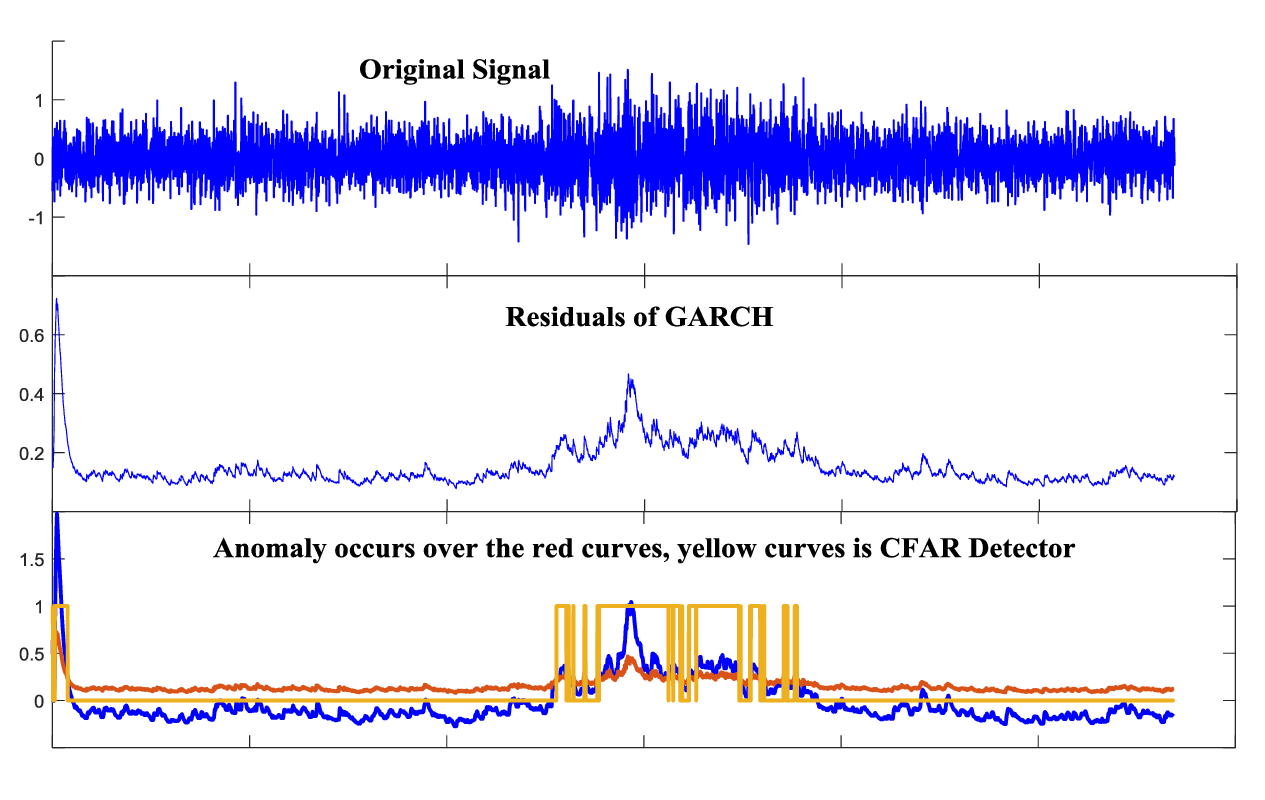}
			\caption{Anomaly in synthesized signal.}
			\label{Fig1}}
	\end{figure}
	\begin{figure}[t]
		\centering{
			\includegraphics[scale=.4]{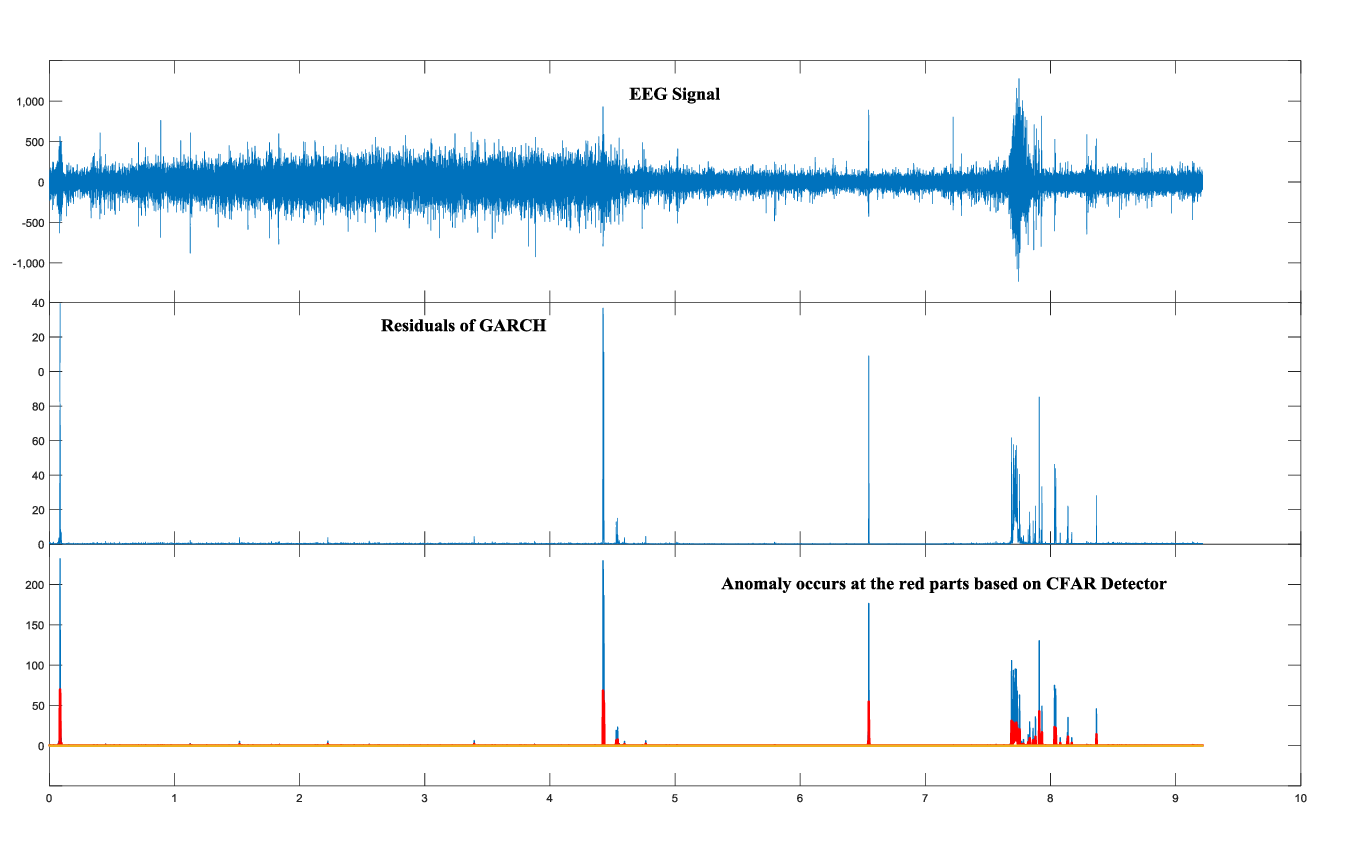}
			\caption{Anomaly in EEG signal of epileptic seizure occurence.}
			\label{Fig2}}
	\end{figure}
	\section{Detection of Simulated and natural anomalies}
\begin{figure}[t]
		\centering{
			\includegraphics[scale=.4]{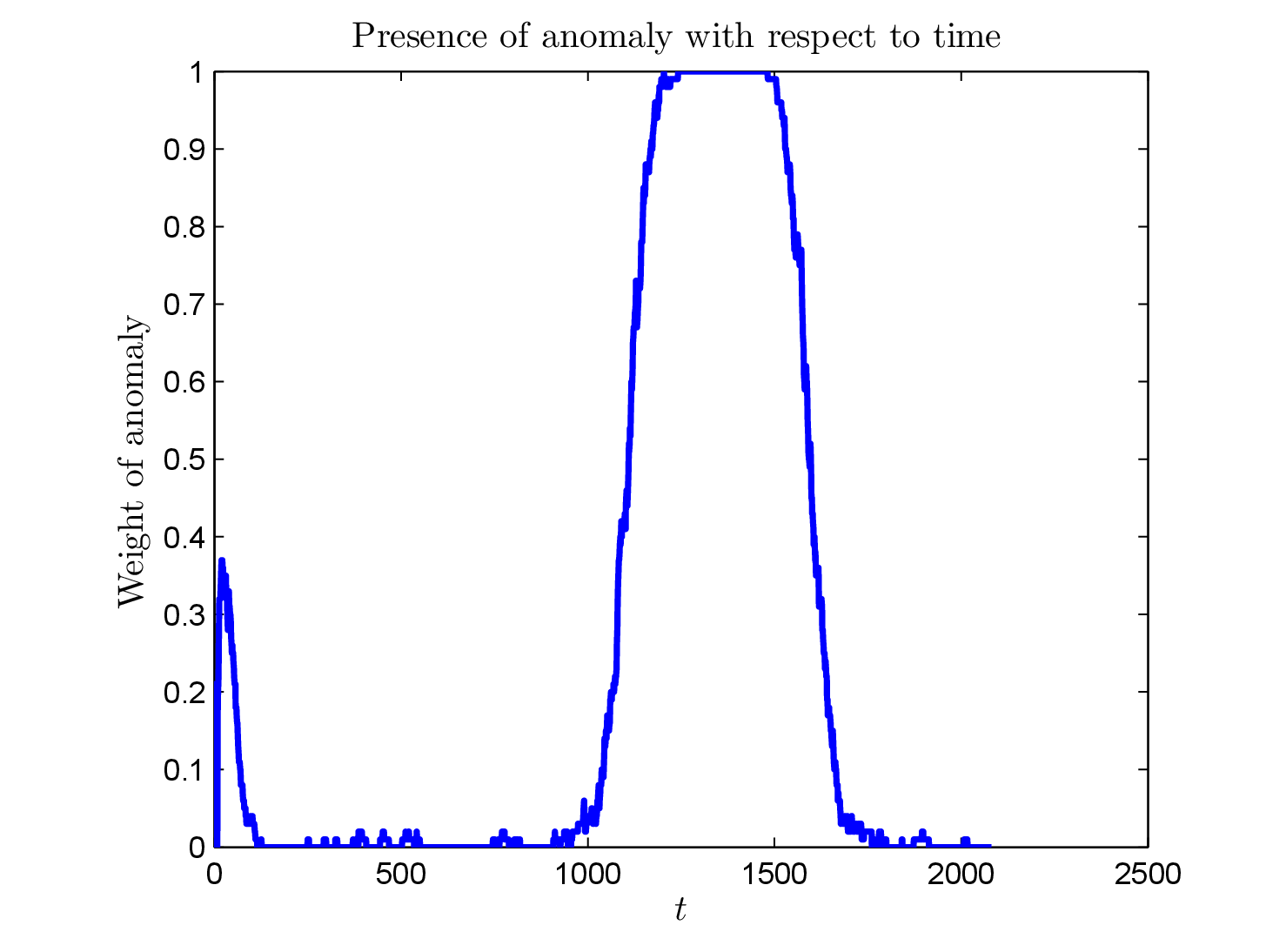}
			\caption{Monte-Carlo simulation of observation to reveal the location in time of the anomalous pattern.}
			\label{Fig5}}
	\end{figure}

\begin{figure}[t]
		\centering{
			\includegraphics[scale=.3]{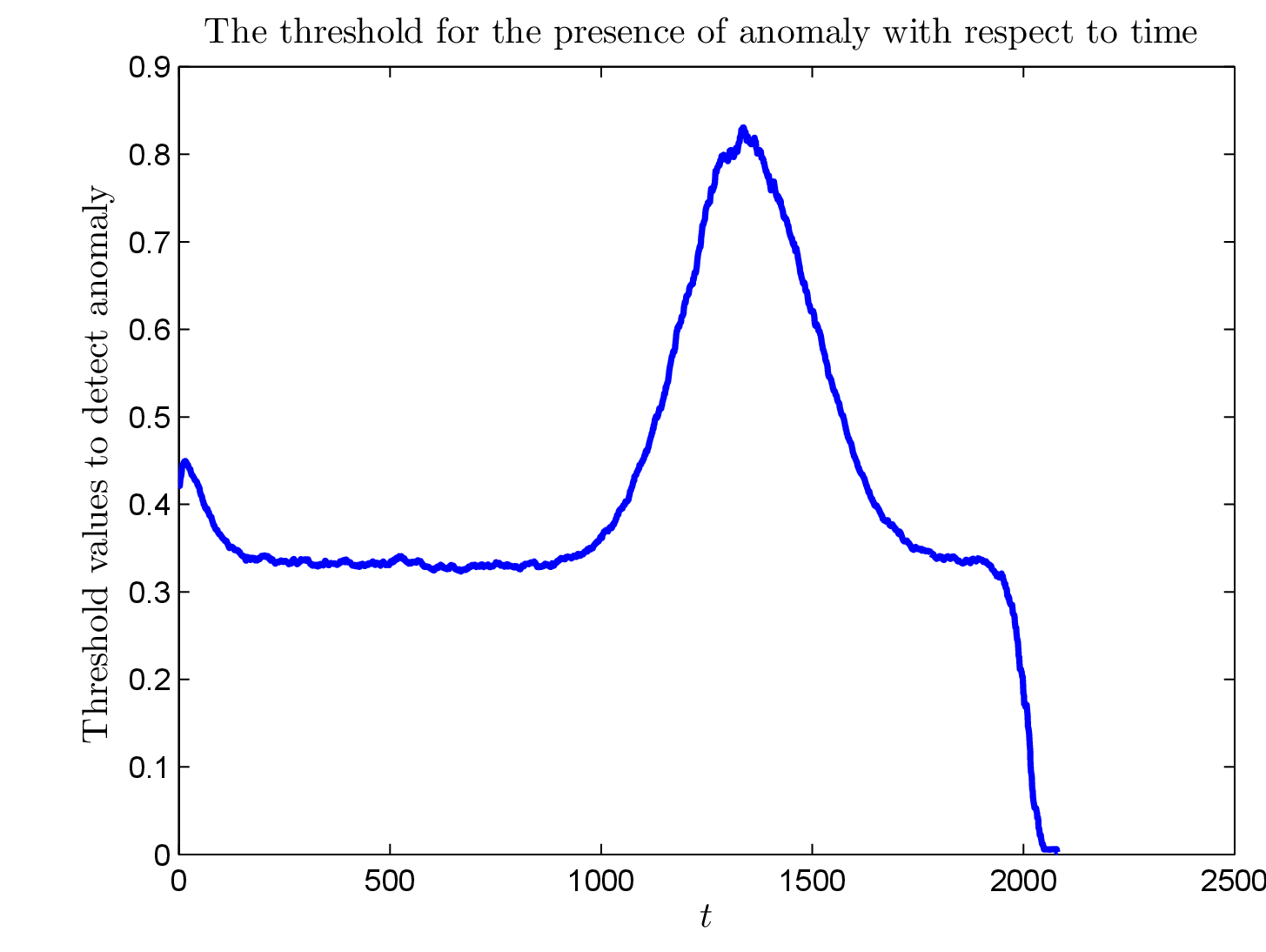}
			\caption{Monte-Carlo simulation of threshold  to reveal the location in time of the anomalous pattern.}
			\label{Fig6}}
	\end{figure}

	For anomalous natural and synthesized signals, this modeling is tested and results seem plausible. First we model a synthesized signal possessing one anomaly pattern which is gradually provoked. As it is seen in Figure~\ref{Fig1}, signal is anomalous, and the model well detects and traces the anomaly pattern. In fact in many natural time series, when anomaly occurs, the variance of time series tends to fluctuate, for instance in epileptic seizures it is more visible. Not only during the seizure attack, but also before seizure attack, powerful patterns of anomaly which affects the constancy of variance, emerge through a discrete occurrence. In the previous work \cite{mohamadi2017arima}, ARIMA-GARCH modeling for epileptic seizure prediction, we have presented ARIMA-GARCH as a tool for seizure prediction, here extended it to any anomalous pattern detection for a wide range of signals including natural signals using CFAR. In Figure~\ref{Fig2} we can see GARCH-related fluctuations are started several minutes before the seizure, which are detected finely using CFAR detector. Actually CFAR detector adaptively sets the threshold and contrasts the fluctuations based on that; so here we it can be seen that any GARCH-related fluctuation could be initially detected by CFAR detector. Considering the original signal and its modeling in case of EEG seizure attack, the interesting thing  is that our modeling do not care even about the powerful fluctuations which are not originated from seizure pattern (such as those fluctuations in the first half-interval of EEG signal in Figure~\ref{Fig2}, it only cares about the fluctuations representing heteroskedastic nature of seizure. In Figure~\ref{Fig5}, the average of 100 run Monte-Carlo simulations is illustrated to depict the presence of anomaly. In this figure, it is seen that the conditional mean, $X_t$ which does not exhibit heteroskedasticity is vanished, but, the anomalous
	signal $Z_t$ remains percussive around the time the simulated anomaly is present, moreover, at the very beginning of signal, though, no anomaly is present, but the standard deviation suddenly changes from zero to some value, an indication of heteroskedasticity, CFAR detects an anomaly for a very short time. Figure~\ref{Fig6} is the average threshold over the 100 runs, again an indication of presence of an anomaly at very beginning of the simulated data is observed, but, a stronger threshold level is estimated around the time where the actual anomaly is present.
	
	\section{Conclusion}
	In this paper, detection of an anomalous signal is declared in presence of a time varying mean. The time varying mean is assumed to follow an ARIMA and the anomalous signal as a GARCH model. The observed signal may possess anomalous pattern for some time. This time varying pattern is captured adaptively using an adaptive CFAR. The proposed method does not require a priori knowledge about the presence or absence of an anomaly, this is a result of hybrid modeling signal of interest using ARIMA-GARCH, if the GARCH component is significant provides a clue for time varying standard deviation, an indication of presence of anomaly in the signal. Furthermore, the time this anomaly occurs in determined by CFAR.
	\bibliographystyle{IEEEbib}
	\bibliography{strings,Birth}
	
\end{document}